# How to explain the sensitivity of DNA double-strand breaks yield to $^{125}$I position?

Mario Enrique Alcocer Ávila[a1] , Elif Hindié[b,c] and Christophe Champion[a]

[a]*Université de Bordeaux-CNRS-CEA, Centre Lasers Intenses et Applications, UMR 5107, 33405 Talence, France*

[b]*Université de Bordeaux, INCIA, CHU de Bordeaux - Service de Médecine Nucléaire, 33604 Pessac, France*

[c]*Institut Universitaire de France (IUF), 75231 Paris Cedex 05, France*

---

[1]Corresponding author, currently at Université Claude Bernard Lyon 1-CNRS-IN2P3, Institut de Physique des deux Infinis, UMR 5822, 69622 Villeurbanne, France

E-mail : mario-enrique.alcocer-avila@univ-lyon1.fr

**Abstract**

**Purpose:** Auger emitters exhibit interesting features due to their emission of a cascade of short-range Auger electrons. Maximum DNA breakage efficacy is achieved when decays occur near DNA. Studies of double-strand breaks (DSBs) yields in plasmids revealed cutoff distances from DNA axis of 10.5-12 Å, beyond which the mechanism of DSBs moves from direct to indirect effects, and the yield decreases rapidly. Some authors suggested that the average energy deposited in a DNA cylinder could explain such cutoffs. We aimed to study this hypothesis in further detail.

**Materials and methods:** Using the Monte Carlo code CELLDOSE, we investigated the influence of the $^{125}$I atom position on energy deposits and absorbed doses per decay not only in a DNA cylinder, but also in individual strands, each modeled as 10 spheres encompassing the fragility sites for phosphodiester bond cleavage.

**Results:** The dose per decay decreased much more rapidly for a sphere in the proximal strand than for the DNA cylinder. For example, when moving the $^{125}$I source from 10.5 Å to 11.5 Å, the average dose to the sphere dropped by 43%, compared to only 13% in the case of the cylinder.

**Conclusions:** Explaining variations in DSBs yields with $^{125}$I position should consider the probability of inducing damage in the proximal strand (nearest to the $^{125}$I atom). The energy received by fragility sites in this strand is highly influenced by the isotropic (4π) emission of $^{125}$I low-energy Auger electrons. The positioning of Auger emitters for targeted radionuclide therapy can be envisioned accordingly.

**Keywords:** Iodine-125; Auger emitters; DNA damage; radiation dosimetry; Monte Carlo simulation

**Introduction**

There is a growing interest in using Auger electron (AE)-emitting radionuclides for targeted radionuclide therapy (Ku et al. 2019, Howell 2020). Because of the emission of a cascade of very short range AE with highly localized energy deposition, a common approach is to design pharmaceuticals able to carry the radionuclides in the proximity to the DNA of tumor cells. Indeed, DNA has historically been considered the main target in radiation therapy (Ku et al. 2019). Moreover, the distance from the decay site to the DNA structure seems of fundamental importance for inducing double-strand breaks (DSBs) and ultimately lethal damage to tumor cells. Iodine-125 ($^{125}$I) is the most extensively investigated AE emitter. $^{125}$I decays to $^{125}$Te by electron capture ($T_{1/2}$ = 59.4 days). The decay of $^{125}$I is accompanied by the emission of an average of 24 AE per decay (International Commission on Radiological Protection 2008, Howell 2020). As shown in Figure 1, the energy of AE ranges from 23 eV to 30.3 keV and that of internal conversion electrons (CE) from 3.7 to 35.5 keV (Figure 1).

*In vivo* studies on mice showed that the radiotoxicity of $^{125}$I depends on the fraction of radionuclide that is incorporated into the cell nucleus, and that the efficacy at cell killing of $^{125}$I is greatly reduced when it is located only in the cytoplasm or outside the cells (Kassis et al. 1987; Rao et al. 1990).

Several theoretical and experimental studies have evaluated the efficacy of radiopharmaceuticals containing $^{125}$I for inducing DSBs. On the theoretical side, Humm and Charlton (1989) performed electron track structure simulations to study the energy deposition of AE sources in a cylindrical DNA model. They studied the probability of DSBs production per $^{125}$I decay at several distances from the DNA axis. They observed that the probability for a $^{125}$I decay to produce a DSB falls rapidly within the first nm separating the decay site from DNA. Various compounds, which position the $^{125}$I atom at different distances from the DNA axis, have been tested (Lobachevsky et al. 2008;

Balagurumoorthy et al. 2012). A steep decrease in the probability of inducing DSBs when the distance to DNA increases was observed in the plasmid DNA breakage experiments of Lobachevsky et al. (2008) and Balagurumoorthy et al. (2012). Besides, in these experiments the radical scavenger dimethyl sulfoxide (DMSO) was used to disentangle the contribution of direct and indirect effects of radiation to the DSBs yields. The results of Lobachevsky et al. (2008) showed that indirect effects predominate beyond a distance of ≈ 11 Å between the $^{125}$I atom and the DNA axis; Balagurumoorthy et al. (2012) proposed the existence of a cutoff at a distance of 12 Å between the $^{125}$I atom and the DNA axis, beyond which DSBs production is entirely due to indirect effects. More recently, Pereira et al. (2017) performed plasmid DNA irradiation studies with $^{125}$I-labelled acridine orange derivatives. From their experiments, Pereira et al. (2017) found a cutoff in the range 10.49–11.04 Å, for moving from direct to indirect effect, which is roughly in agreement with the value proposed by Lobachevsky et al. (2008) and Balagurumoorthy et al. (2012).

In the search for an explanation of their results, Pereira et al. (2017) raised the hypothesis that the cutoff could be explained by studying the energy deposited in a cylinder of nanometric dimensions (2.3 nm of diameter and 3.4 nm of height) representing a DNA segment of 10 base pairs. They computed the deposited energy in the cylinder by means of Monte Carlo (MC) simulations with the MCNP6 code and tried to correlate their calculations with the DSBs yields measured in their plasmid DNA experiments. According to Pereira et al. (2017), the energy deposited in the cylinder shows a steep variation with the distance to the DNA axis, suggesting the existence of a critical distance beyond which the direct effects stop being effective.

Because DSBs and complex DNA damage involve energy deposit in the two strands of DNA, the aim of this work was to analyze the influence of $^{125}$I position on the energy deposits and absorbed dose by considering, not only the average energy deposit in a cylindrical DNA segment, but also a finer description taking into account the fragility sites of phosphodiester bonds in each DNA strand.

**Materials and methods**

MC track structure simulations were performed with the CELLDOSE code (Champion et al. 2008) to evaluate the energy deposited by a point source of $^{125}$I in: (1) a cylinder of height = 34 Å and radius = 11.85 Å, representing a DNA segment of 10 base pairs; (2) spheres of radius 1.7 Å in contact with each other along each DNA strand (10 spheres per strand).

All the volumes considered in the simulations were assumed to contain unit density water. Each CELLDOSE simulation consisted of $10^8$ decays of $^{125}$I. The electron energies were sampled from the spectrum of $^{125}$I provided by the ICRP Publication 107 (International Commission on Radiological Protection 2008), as shown in Figure 1 (a second energy spectrum of $^{125}$I, taken from Lee et al. (2016), was used for comparison purposes). The electrons emitted by $^{125}$I were followed down to an energy of 7.4 eV, and the residual energy was considered as locally deposited.

In our simplified model, the two strands of DNA were considered as unfolded and on the same plane, as illustrated in Figure 2.

The length of each strand was taken as 34 Å (equal to 10 base pairs) and the two strands were separated by 23.7 Å, the diameter of DNA (Raisali et al. 2013). The distance between the DNA axis and the center of any of the spheres was of 10.15 Å, and that to the outer border of the spheres 11.85 Å (Figure 2). The volume of each target sphere positioned along a DNA strand was assumed to encompass one of the fragility sites of a phosphodiester bond (Zheng et al. 2005), as depicted in Figure 3.

In its initial position, the radioactive source of $^{125}$I was placed in the middle of the DNA central axis. The source position was subsequently shifted by steps of 0.5 Å toward the virtual border of the DNA cylinder, then leaving the cylinder. During this shift the $^{125}$I source gets closer to the proximal DNA strand, before moving away again. Figure 2 shows that, when considering the two spheres in the middle of the proximal strand (S1 and S1'), the solid angle $\Omega$ subtended by these spheres increases as the source approaches the proximal strand, reaching a maximum for a distance of 10.15 Å from DNA axis. Conversely, $\Omega$ decreases when $d > 10.15$ Å because the source moves away from the spheres. As regards the opposite strand, the shift will progressively increase the distance from the $^{125}$I source.

For each source configuration, a MC simulation was performed to compute the energy deposited in the cylinder and the spheres. The energy deposits scored in the targets were then converted to absorbed doses per decay. Because of the symmetry of the system (Figure 2), the energy deposited in the spheres S1-S5 was equal to the energy deposited in spheres S1'-S5', respectively. Again, when considering the opposite strand, the energy deposited in spheres S6-S10 was equal to the energy deposited in spheres S6'-S10', respectively.

**Results**

Figure 4 shows our findings with MC CELLDOSE simulations of the energy deposited per decay in the cylinder representing a DNA segment of 10 base pairs. The solid line corresponds to the simulation using the $^{125}$I spectrum of the ICRP Publication 107 (International Commission on Radiological Protection 2008) and the dashed line corresponds to a simulation using the $^{125}$I spectrum proposed by Lee et al. (2016). The results reported by Pereira et al. (2017) (dash-dotted line) are shown for comparison.

It can be seen from Figure 4 that the energy deposited in the cylinder gradually decreases as the distance from the $^{125}$I source to the DNA axis increases. The decrease is more pronounced as the source reaches the border of the cylinder and moves further away (inflection point at 11.85 Å and 11.5 Å for this work and Pereira et al. (2017), respectively). In all cases, the energy deposited by decay drops roughly by about 30% when the distance between the $^{125}$I source and the DNA axis increases by 1.5 Å.

Overall, it can be seen from Figure 4 that, despite discrepancies in the specific energy deposition values (arising from the use of different spectra and different MC codes), the general behavior of all curves is essentially the same regardless of the spectrum, i.e., a monotonous decrease in energy deposition with the distance to the DNA axis.

Figure 5 shows the absorbed dose per decay in the target spheres S1, S2 and S6 of the DNA model depicted in Figure 2, and in the cylinder representing a DNA segment of 10 base pairs (height = 34 Å and radius = 11.85 Å).

Similarly to Figure 4, it can be observed that as the radiation source leaves and moves away from the DNA volume, there is a progressive decrease in the absorbed dose per decay to the cylinder, with a somewhat more pronounced decrease at the point where the source leaves the cylinder (11.85 Å). Likewise, a monotonous decrease in the absorbed dose per decay is also observed for the sphere S6, located in the distal (opposite) strand. In contrast, the dose to the sphere S1 presents a behavior consistent with the one observed in Figure 2 for the solid angle, i.e. the absorbed dose per decay increases as the source approaches the target sphere, reaching a maximum between 10 Å and 10.5 Å, and then decreases rapidly as the source moves away from the sphere. The results for the sphere S2 show a similar but less pronounced trend.

It is thus apparent that the dose to components of the proximal strand, closest to the radiation source, here represented by the sphere S1, is much more sensitive to $^{125}$I position than the dose to the cylinder. To better appreciate this fact, we computed the dose ratio taking the value at 10.5 Å as reference (the theoretical maximum dose would be found for 10.15 Å as apparent in Figure 2). The results are shown in Table 1.

We have highlighted in Table 1 the doses and dose ratios at different cutoff values. For example, taking into account experimental results of Pereira et al. (2017), showing a cutoff between 10.5 Å – 11 Å, Table 1 shows that as the source moves from 10.5 Å to 11 Å from DNA axis, the dose to the sphere S1 located in the proximal strand decreases by 22%, while the decrease in the average dose to the cylinder is only 6%. Likewise, taking into account the cutoff of 12 Å identified by Balagurumoorthy et al. (2012), the dose to the sphere S1 decreases by 58%, when moving from 10.5 Å to 12 Å, while the same shift reduces the average dose to the cylinder by 24%.

**Discussion**

The success of radiation therapy with AE emitters strongly depends on the ability to place the radioactive atom near the DNA structure. Multiple studies have shown that the DSBs yield is highly dependent on the distance between an AE emitter and DNA. These studies identified cutoff values of about 10.5-12 Å (distance between radionuclide position and DNA central axis), beyond which the yield decreases rapidly and a switch occurs from direct to indirect effect. Pereira et al. (2017) suggested that the cutoff can be explained by abrupt decrease in energy deposited in the whole DNA cylinder. We here suggest that an explanation to the cutoff needs to take into account the energy deposited in individual strands of DNA rather than in the DNA as a whole.

We introduced an alternative description of a DNA segment in which we considered in more detail the energy deposited in spheres encompassing the fragility sites of phosphodiester bonds in each DNA strand. We found that the dose to the proximal strand represented by the sphere closest to the $^{125}$I source (S1; Figure 2) is very sensitive to distance and experiences a much steeper decrease than the one observed in the whole cylinder. For example, Table 1 shows that the dose to the sphere S1 decreases by 43%, when moving from 10.5 Å to 11.5 Å, while the same shift reduces the average dose to the cylinder by 13%. For the same variation in distance, there is only a 10% decrease in the energy deposited in the sphere S6 on the opposite strand. Overall, the AE emitter would be ideally positioned closest to one of the strands and towards the interior of the DNA cylinder, in order to also keep the dose to the opposite strand as high as possible.

A DNA DSB is the break of the sugar–phosphate backbone on both strands and at sites located directly opposite each other or just a few nucleotides apart (up to ~10 base pairs). It should be noted that several SSBs in one DNA strand do not constitute a DSB. The reason why the proximity of $^{125}$I to one of the strands will finally result in more DSBs can be explained, however, by the high sensitivity of the proximal strand to the position of the $^{125}$I source. Indeed, the energy deposited, and therefore the number of SSBs, increases steeply when $^{125}$I is closest to the proximal strand. A higher induction of SSBs on different adjacent sites along the proximal strand will have the effect of increasing as well the probability of producing a DSB in the opposite strand, even when the energy deposited in the latter is considerably less. Let us examine for instance the behavior of the total energy deposited in the central spheres of both strands (represented by S1 on one strand and S6 on the opposite strand, see Figure 2) when $^{125}$I is at y = 0 Å (i.e. the source is at the center of the DNA cylinder on the DNA axis), and when y = 10 Å (i.e. source closest to S1). In the former case (y = 0 Å), the total energy per decay is equal to S1 + S6 = 0.34 eV + 0.35 eV ≈ 0.7 eV. However, at y = 10 Å, we have S1 + S6 = 15 eV + 0.1 eV = 15.1 eV. In relative terms, the total energy deposit has increased by a factor ≈ 22. Again, if we consider the sum of the energy deposited in all 20 spheres (10 on each strand, Figure 2), the sum is equal to 4.2 eV when $^{125}$I is at the center of the cylinder, and 35.3 eV (34.7 eV for the 10 spheres on the proximal strand + 0.6 eV for the 10 spheres on the opposite strand) when $^{125}$I is at 10 Å from the center and close to S1. We see that in the latter case the total energy deposit is more than 8 times higher than when the source is at the center. It is clear from these examples that the global energy deposited at fragility sites on the two strands of DNA is enhanced when $^{125}$I atoms are located very close to a DNA strand. The probability of inducing a DSB will be globally enhanced as well because of the large increase in the number of SSBs occurring in the fragility sites close to $^{125}$I, despite the fact that the occurrence of SSBs on the opposite strand is expected to be slightly lower. In conclusion, the net result in terms of DSBs yields will be favored by the closest proximity of the $^{125}$I source to one of the strands.

When considering the influence of distance on the energy deposited in the cylinder, our results are in qualitative agreement with those of Pereira et al. 2017 (Figure 4), that is the curves relating energy deposited to distance show similar shapes but with different absolute values. Please note that such differences can be partly explained by the use of different emission spectra for $^{125}$I and different MC codes.

Pereira et al. (2017) performed their simulations with the MCNP6 code using the $^{125}$I spectrum provided by Howell (1992), while we used the spectrum of $^{125}$I taken from the ICRP Publication 107 (International Commission on Radiological Protection 2008). For further verification, we used an alternative $^{125}$I spectrum recently reported by Lee et al. (2016), and the results of this simulation is also presented in Figure 4.

While we investigated in finer details the pattern of energy deposits from $^{125}$I, our study has limitations. A full explanation of the impact of AE emitters would require taking into account additional phenomena, such as fragility sites other than phosphodiester bonds (e.g., *N*-glycosidic bonds) (Zheng et al. 2005), molecular fragmentation resulting from positive charge build-up during the AE cascade emission, also termed "Coulomb explosion" (Pomplun and Sutmann 2004), electron transfer phenomena (for example to the phosphodiester bond (Zheng et al. 2005)), etc. It would be expected, however, that these local phenomena would also involve the strand most proximal to $^{125}$I decay site. Finally, let us recall that our MC code follows electrons down to an energy of 7.4 eV, after which residual energy is considered locally absorbed. Therefore there is also an uncertainty associated to the energy deposits of subcutoff electrons. In our model, the contribution of these subcutoff electrons to the calculated dose does not exceed 15%. Supplementary Figure 1 shows the current simulation considering the very low-energy electrons as locally absorbed and an alternative simulation not taking into account the residual energy carried by these electrons.

**Conclusions**

Previous studies have shown the sensitivity of the DSBs yield according to the position of an AE emitter with respect to DNA axis. The results obtained with the DNA model considered in this work reveal a steep decrease of the dose taking place within the first angstroms separating the $^{125}$I source from the proximal DNA strand. In contrast, the simulations based on a simpler model, i.e. a cylindrical volume representing a DNA segment of 10 base pairs, showed a more monotonous decrease of the dose. Thus, any analysis of the influence of an AE emitter's position on DSBs yield should first consider the energy deposited near the fragility sites of phosphodiester bonds and the probability of inducing a SSB on the proximal DNA strand, taking into account the multiple low-energy electrons characteristic of the AE emission process.

**Acknowledgments**

MEAA acknowledges the financial support of the Site de Recherche Intégrée sur le Cancer de Bordeaux, Bordeaux Recherche Intégrée en Oncologie (SIRIC BRIO).

**Disclosure statement**

The authors report no conflicts of interest.

**Notes on contributors**

Mario E. Alcocer Ávila earned a PhD in physics from Bordeaux University. He is currently a postdoctoral researcher at the University of Lyon. His research interests include radiation and computational physics.

Elif Hindié is Professor of Nuclear Medicine at Bordeaux University and Hospitals. He is a specialist in the use of radionuclides for imaging and therapy in oncology as well as in endocrine diseases.

Christophe Champion is Professor of Physics at Bordeaux University. His research activities consist in describing in detail the collisional mechanisms induced by charged particles in biological media. In this context, he developed various theoretical models to model the elastic and inelastic processes, all described within the quantum mechanical framework. In parallel, C. Champion has developed a series of Monte Carlo codes for describing the charged particle track structure, namely, EPOTRAN (an acronym for Electron and POsitron TRANsport in biological medium) devoted to electron and positron transport - with its extension CELLDOSE focused on microdosimetry of radionuclide emitters -, and TILDA as well as its recent extension TILDA-V, devoted to ion transport in liquid water and DNA components, respectively.

**ORCID**

Mario Enrique Alcocer Ávila  https://orcid.org/0000-0003-3999-552X

Elif Hindié  https://orcid.org/0000-0003-2101-5626

Christophe Champion  https://orcid.org/0000-0002-6202-3465

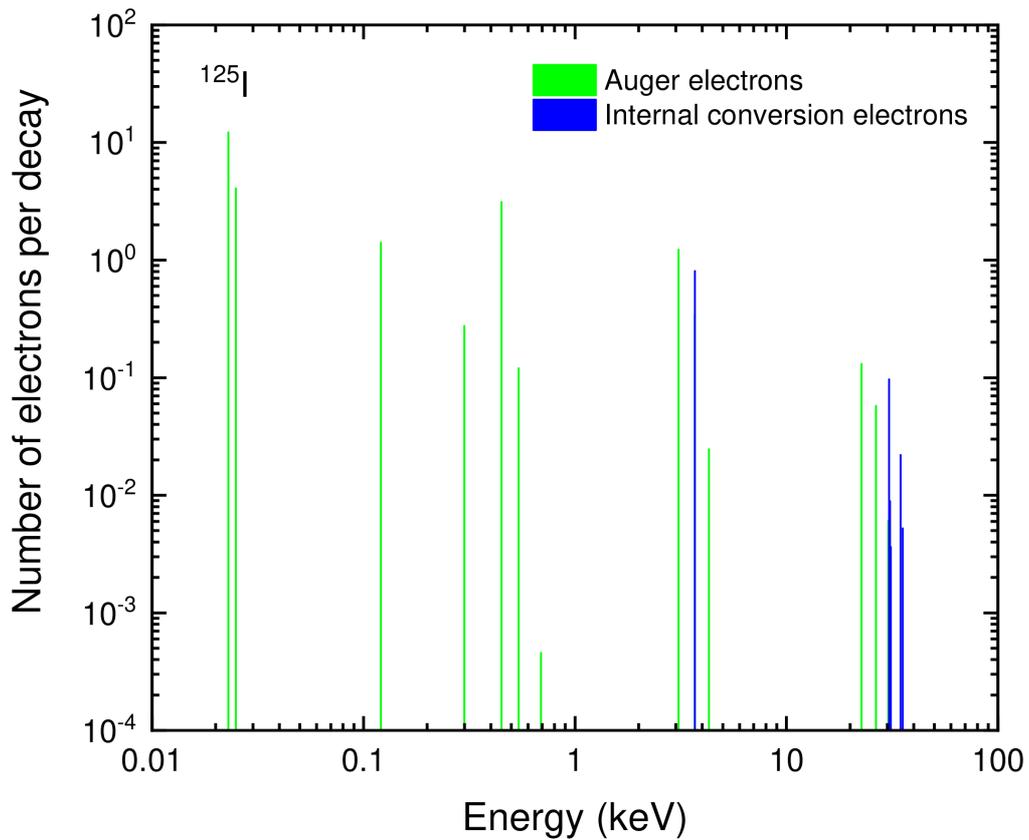

Figure 1. Spectrum of $^{125}$I (International Commission on Radiological Protection 2008).

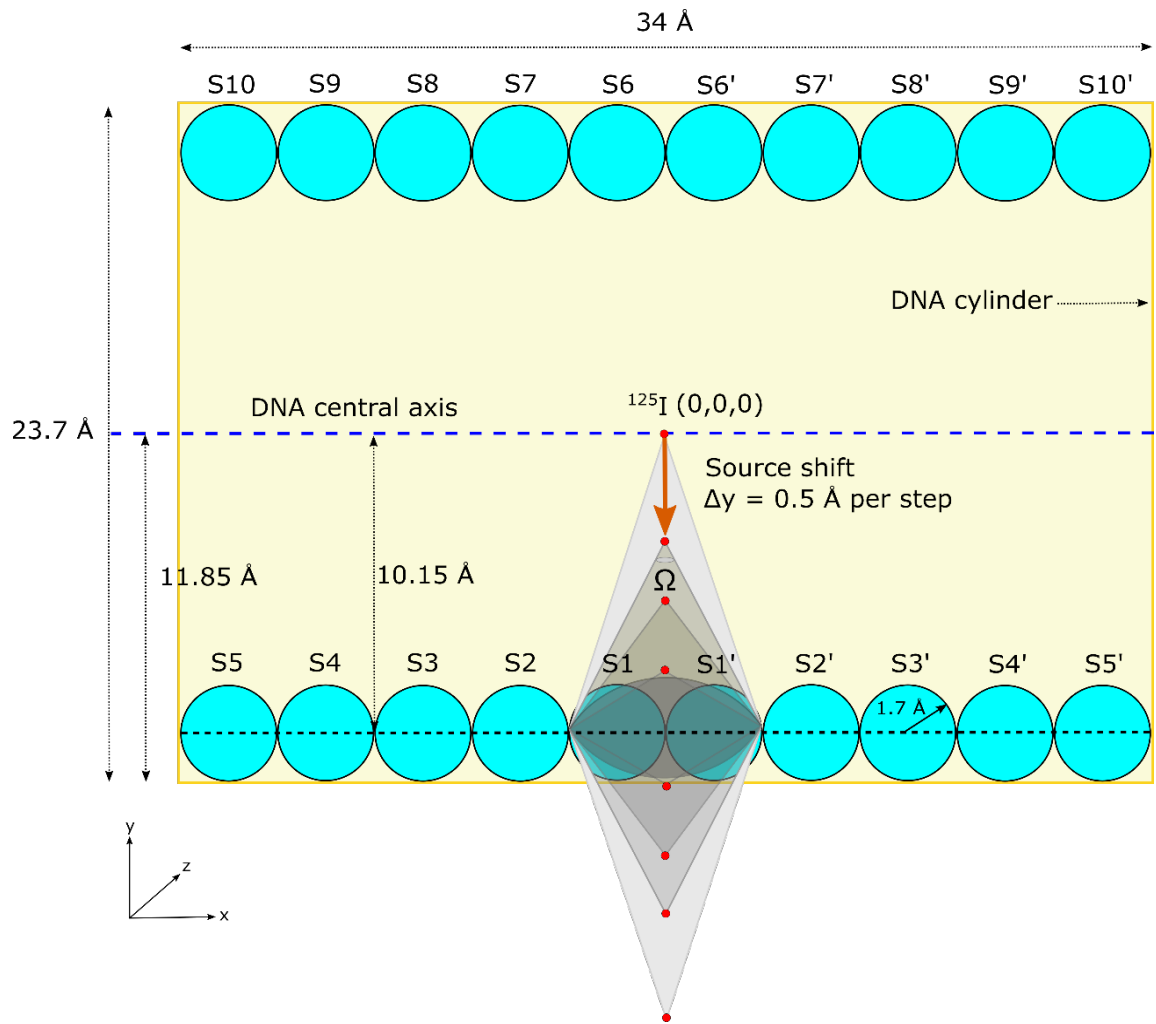

Figure 2. Illustration of the DNA geometry simulated with CELLDOSE. The area depicted in yellow represents a lateral view of the DNA cylinder.

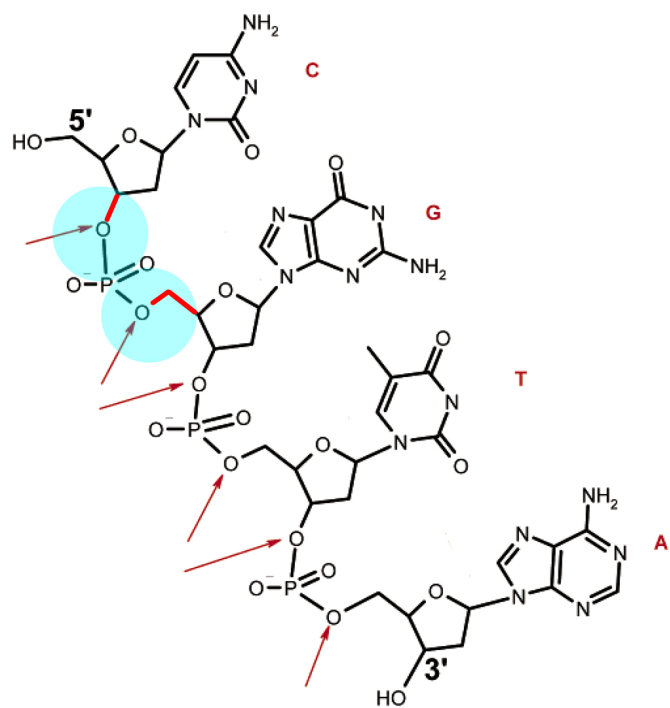

Figure 3. Correspondence between fragility sites in phosphodiester bonds and the spheres in our DNA model (circled regions). Note that two consecutive spheres encompass different fragility sites of the phosphodiester bond. Adapted with permission from Zheng et al. Copyright (2005) American Chemical Society.

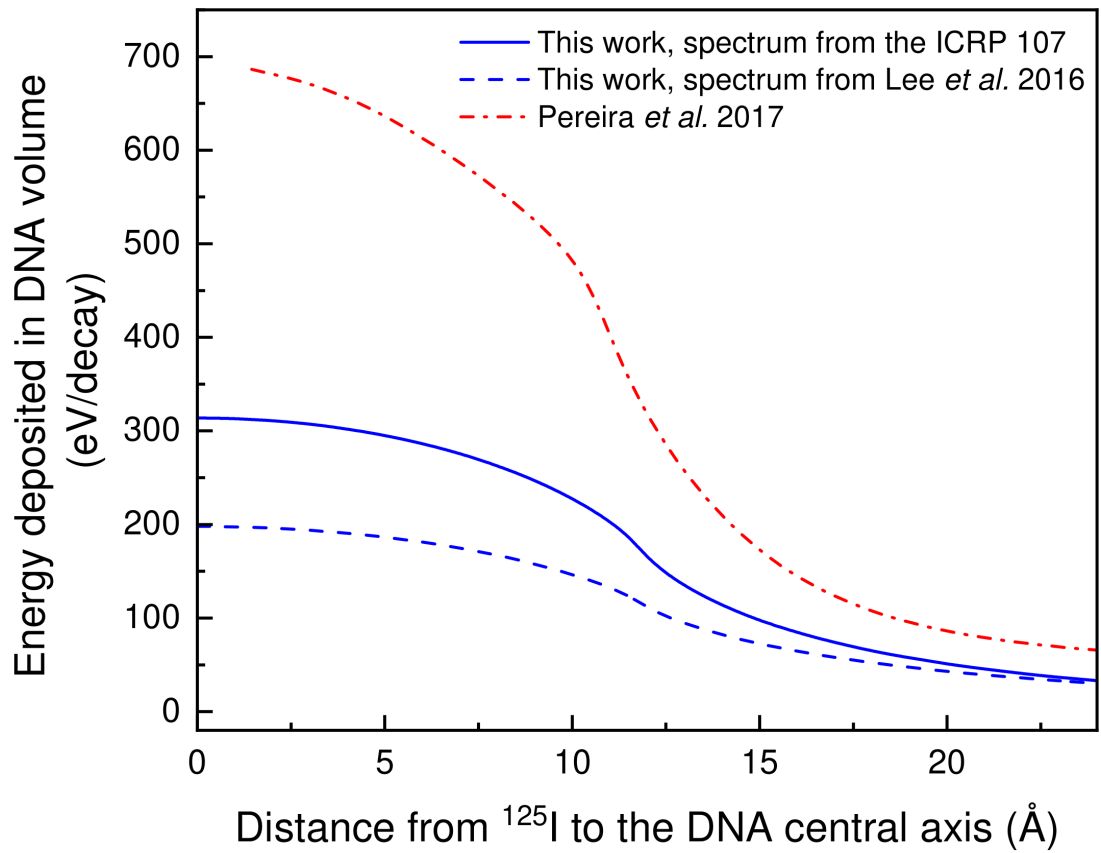

Figure 4. Energy deposited per decay in the cylinder representing a DNA segment of 10 base pairs.

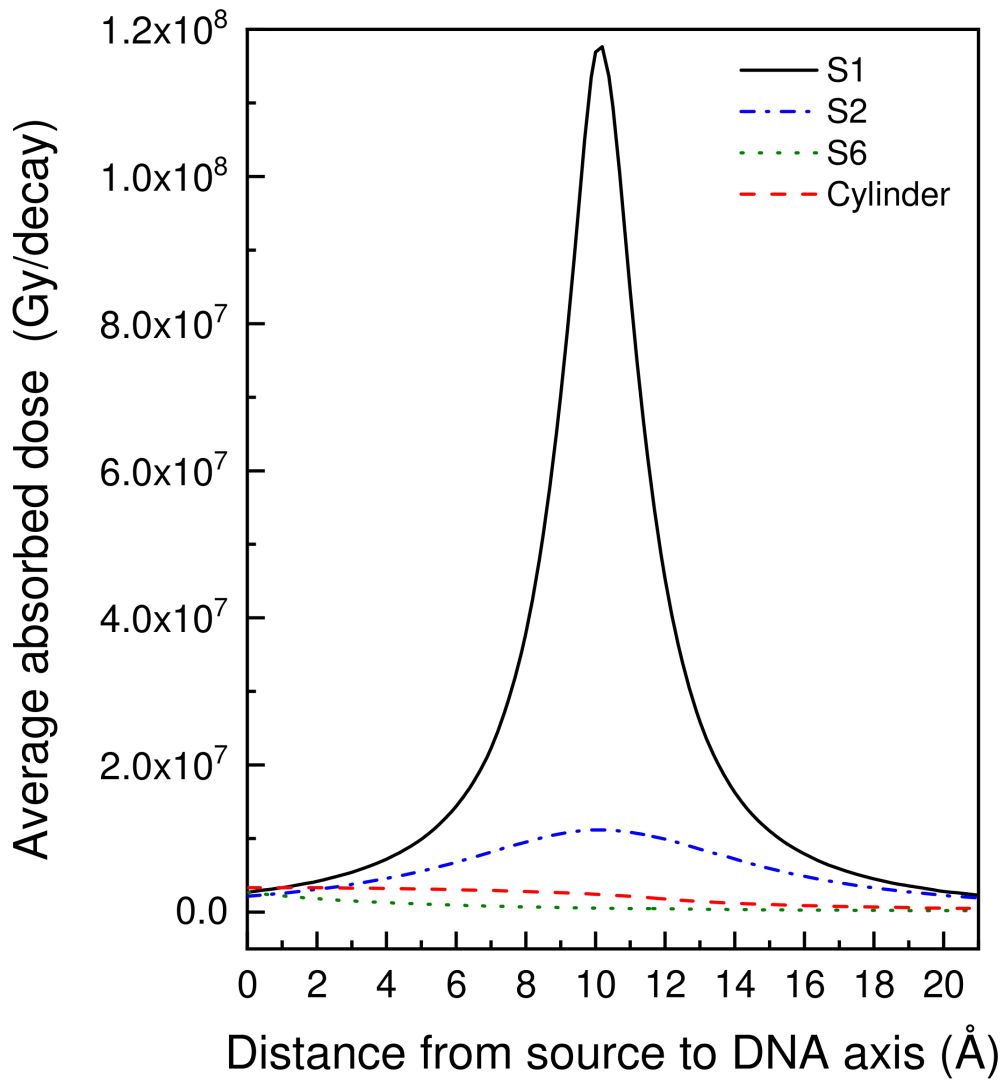

Figure 5. Average absorbed dose per decay in the target spheres S1 (solid line), S2 (dash-dotted line) in the proximal strand and S6 (dotted line) in the opposite strand of the DNA model represented in Figure 2. The average dose to the cylinder is shown as well (dashed line). All simulations used the $^{125}$I spectrum of the ICRP Publication 107.

Table 1. Absorbed dose per decay in the spheres S1 and S6, and in a cylinder representing a DNA segment of 10 base pairs, as a function of the source distance to the DNA central axis. The dose ratio taking the dose at 10.5 Å as reference is shown as well.

| Source distance to DNA axis (x in Å) | Dose S1 (Gy/decay) | S1(x)/ S1(10.5) | Dose S6 (Gy/decay) | S6(x)/ S6(10.5) | Dose cylinder (Gy decay) | Cyl(x)/ Cyl(10.5) |
|---|---|---|---|---|---|---|
| 0 | 2.68E+06 | 0.03 | 2.69E+06 | 5.2 | 3.34E+06 | 1.5 |
| 1 | 3.33E+06 | 0.03 | 2.20E+06 | 4.3 | 3.33E+06 | 1.4 |
| 2 | 4.18E+06 | 0.04 | 1.81E+06 | 3.5 | 3.31E+06 | 1.4 |
| 3 | 5.41E+06 | 0.05 | 1.52E+06 | 2.9 | 3.27E+06 | 1.4 |
| 4 | 7.18E+06 | 0.07 | 1.28E+06 | 2.5 | 3.21E+06 | 1.4 |
| 5 | 9.91E+06 | 0.09 | 1.10E+06 | 2.1 | 3.14E+06 | 1.4 |
| 6 | 1.44E+07 | 0.13 | 9.52E+05 | 1.8 | 3.05E+06 | 1.3 |
| 7 | 2.23E+07 | 0.21 | 8.22E+05 | 1.6 | 2.93E+06 | 1.3 |
| 7.5 | 2.88E+07 | 0.26 | 7.60E+05 | 1.5 | 2.87E+06 | 1.2 |
| 8 | 3.79E+07 | 0.35 | 7.16E+05 | 1.4 | 2.79E+06 | 1.2 |
| 8.5 | 5.13E+07 | 0.47 | 6.73E+05 | 1.3 | 2.71E+06 | 1.2 |
| 9 | 7.03E+07 | 0.65 | 6.30E+05 | 1.2 | 2.63E+06 | 1.1 |
| 9.5 | 9.47E+07 | 0.87 | 5.94E+05 | 1.1 | 2.53E+06 | 1.1 |
| 10 | 1.17E+08 | 1.1 | 5.56E+05 | 1.1 | 2.42E+06 | 1.1 |
| 10.5 | 1.09E+08 | 1.0 | 5.17E+05 | 1.0 | 2.30E+06 | 1.0 |
| 11 | 8.45E+07 | 0.78 | 4.87E+05 | 0.94 | 2.16E+06 | 0.94 |
| 11.5 | 6.18E+07 | 0.57 | 4.66E+05 | 0.90 | 1.99E+06 | 0.87 |
| 12 | 4.53E+07 | 0.42 | 4.37E+05 | 0.85 | 1.75E+06 | 0.76 |
| 12.5 | 3.39E+07 | 0.31 | 4.11E+05 | 0.79 | 1.57E+06 | 0.68 |
| 13 | 2.59E+07 | 0.24 | 3.95E+05 | 0.76 | 1.43E+06 | 0.62 |
| 13.5 | 2.03E+07 | 0.19 | 3.68E+05 | 0.71 | 1.31E+06 | 0.57 |
| 14 | 1.63E+07 | 0.15 | 3.46E+05 | 0.67 | 1.20E+06 | 0.52 |
| 15 | 1.10E+07 | 0.10 | 3.16E+05 | 0.61 | 1.03E+06 | 0.45 |
| 16 | 7.92E+06 | 0.07 | 2.87E+05 | 0.56 | 8.92E+05 | 0.39 |
| 17 | 5.88E+06 | 0.05 | 2.61E+05 | 0.50 | 7.78E+05 | 0.34 |
| 18 | 4.52E+06 | 0.04 | 2.32E+05 | 0.45 | 6.82E+05 | 0.30 |
| 19 | 3.56E+06 | 0.03 | 2.13E+05 | 0.41 | 6.02E+05 | 0.26 |
| 20 | 2.83E+06 | 0.03 | 1.98E+05 | 0.38 | 5.33E+05 | 0.23 |